# Fe doped Magnetic Nanodiamonds made by Ion Implantation

ChienHsu Chen[a], I.C. Cho[b], Hui-Shan Jian[c] and H. Niu[a*]

[a]Nuclear Science and Technology Development Center, National Tsing Hua University, HsinChu 30013, Taiwan

[b]Institute for Radiology Research, Chang Gung University and Chang Gung Memorial Hospital, Taoyuan 33302, Taiwan

[c]Department of Biomedical Engineering and Environment Science, National Tsing Hua University, HsinChu 30013, Taiwan

*Here we present a simple physical method to produce magnetic nanodiamonds (NDs) using high dose Fe ion-implantation. The Fe atoms are distributed inside the NDs without affecting their crystal structure. So the NDs can be still functionalized through surface modification for targeted chemotherapy and the added magnetic property will make the NDs suitable for localized thermal treatment for cancer cells without the toxicity from the Fe atoms being directly in contact with the living tissue.*

Nanoparticle mediated cancer treatments have been widely studied in recent years. The applications can be generally divided into two categories. One is for drug delivery through functionalized nanoparticles by surface modification. The other is for localized radiation therapy such as radio frequency (RF) thermal therapy and boron neutron capture therapy (BNCT) [1; 2]. The former requires the particles having a chemically active surface so they can be functionalized to carry the needed chemicals. The later requires the particles to have certain physical properties so that they can respond to electro-magnetic (EM) waves or other radiations. No matter which application, however, biocompatibility is the foremost requirement. Nanodiamonds (NDs) have been shown to be an ideal nontoxic agent to perform such functions. Various functionalized NDs have been demonstrated to be effective in delivering drugs to cancer cells. NDs, however, are physically very inert and stable [3-5]. They are not magnetic to be able to respond to the EM field. They also cannot be easily excited by particle irradiation to induce nuclear reactions to be useful for localized cancer treatment. Other candidates such as nanoparticles of iron compounds, which can be easily excited by EM field to perform thermal therapy, and boron compounds, which have a very large neutron capture cross section to perform BNCT, all have shown some level of cytotoxicity, and are not perfectly biocompatible.

Therefore, the future success of using boron or magnetic materials for radiation therapy depends greatly on our ability to find the proper agents to carry these materials to the target of interest [6]. Since NDs are excellent drug delivery agent, it will be ideal if we can add additional physical properties so that they can also function as an agent for the radiation therapy mentioned above.

In this brief communication, we report for the first time Fe doped magnetic NDs using the technique of ion implantation [7]. Ion implantation, widely used in semiconductor device fabrication, is a power tool for adding impurities to a host material. It has been previously reported that proton implantation can be used to make NDs florescent due to defects generated by the implantation process [8]. The damage to the crystal lattice can be easily repaired by proper post implant thermal annealing. In our study, the crystal structure of NDs is maintained while they become magnetic because of the added Fe atoms.

The procedure for Fe ion implantation into NDs is shown in figure 1. ND powder (100 nm, Element Six Co.) was first dissolved in deionized water. The solution was then dropped on an oxidized silicon wafer and dried under a lamp. The Si sample covered by NDs was put in a vacuum chamber for ion implantation. Fe ions with an energy of 72 KeV and a dosage of $5\times10^{15}$ a.t./cm$^2$ were implanted into NDs. After implantation the samples were annealed at 600 °C for 3 hours in ambient atmosphere. The attached NDs were then removed from the Si substrate using an ultrasonic bath of deionized water. Because the NDs were stacked on the Si wafer with probably many layers, only a part of the NDs were implanted by Fe ions. The ones that received Fe ions to become magnetic have to be sorted out using a magnetic filter. We used a magnetic-activated cell sorting tool (MACS® Separator, Miltenyi Biotec Co.) as the filter. The Fe doped NDs were attracted by the magnet filter and attached to the adjacent walls when the ND solution passed through the filter. Then the attached magnetic NDs were flushed out by injected deionized water. Finally the magnetic NDs were collected and stored in a beaker with deionized water.

The ion implantation unavoidable introduces damage to the crystal lattice that receives the implantation. As mentioned above an annealing process was used to repair the damage of the NDs. Raman spectroscopy was used to examine the crystal property of NDs before and after implantation. Figure 2 shows the measured Raman spectra of the original NDs, the NDs after Fe implantation and the NDs after implant and annealing. The excitation source was a 532nm green laser. The original NDs had a diamond peak at 1332.5 cm$^{-1}$ and a broad peak at 1579 cm$^{-1}$, which is probably due to surface graphitic structure. The full width at half maximum (FWHM) of the diamond peak was 10.1 cm$^{-1}$. After Fe ion implantation, the diamond peak shifted to 1317 cm$^{-1}$ with a much wider FWHM of 98.5 cm$^{-1}$. The blue shift of the Raman peak and the widening of the FWHM indicate clearly that the crystal lattice was damaged. However after annealing, the diamond peak became very strong and restored to the original position of 1332.4 cm$^{-1}$ and the FWHM was reduced to 11.1 cm$^{-1}$. The other peak related to the surface sp2 bonds, however, disappeared. This clearly indicates that the diamond structure is restored and the unwanted surface graphitic structure or its compounds were cleared up.

We also used transmission electron microcopy (TEM) to exam the NDs implanted with Fe after annealing and the result is shown in figure 3. The aggregated NDs image is shown in Fig. 3a and that of a high resolution lattice image of a single ND is shown in Fig. 3b. No defects were found in the structure. The Energy-dispersive X-ray spectroscopy (EDX) was used to look at the content of various elements in the ND. The spectrum of an original ND is shown in Fig. 3d. The spectrum of the ND with the image shown in Fig 3b is shown in Fig. 3c. While before Fe implantation, no Fe signal was found, both K-alpha and K-beta signals of Fe are clearly seen after implant and annealing. Other elements shown in Fig. 3d are probably from surface contamination of the original NDs. But most of them went away after annealing. The Cu signal was from the sample holder and the Si signal was the Si impurities in the NDs. It was estimated that the Fe content in the ND is around 0.2% in weight. It was because of the presence of these Fe atoms, the NDs became magnetic.

The magnetic NDs can be used in various cancer therapies when we need to control the nanoparticles with the external magnetic field and electro-magnetic radiation. RF heating is one of the examples. Other applications include enhanced magnetic resonance imaging (MRI) and guiding and localization of the NDs to the needed area for cancer treatment. Although only Fe atoms were implanted in NDs in this work, the technique certainly can includes other species that may give NDs with other properties. Adding Boron atoms to NDs by ion implantation will be of particular interest because of the application in BNCT. The technique presently here opens a way for combining the drug delivery capability and the radiation therapy by the same NDs as the carrier agents for cancer treatment.

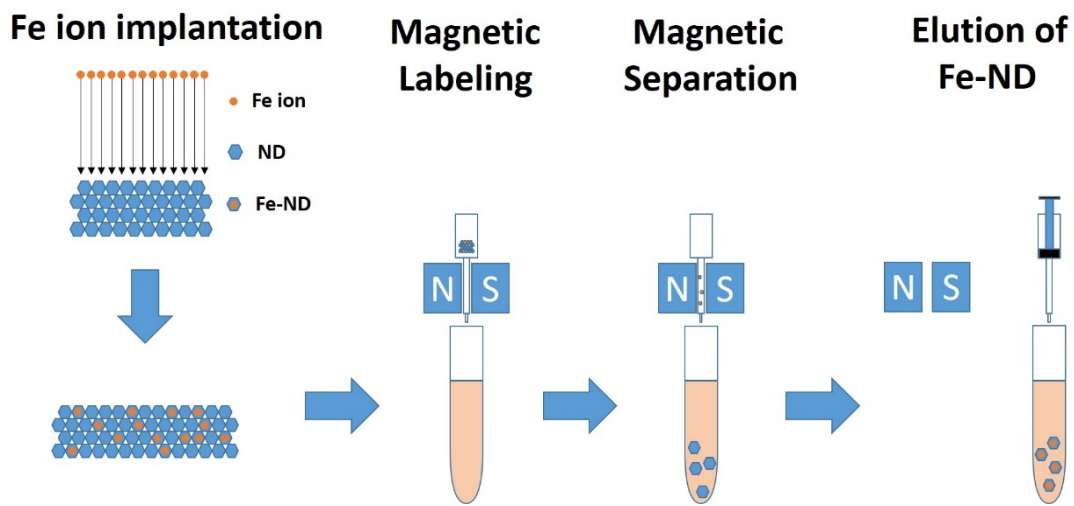

Figure 1. The procedure for Fe ion implantation into NDs.

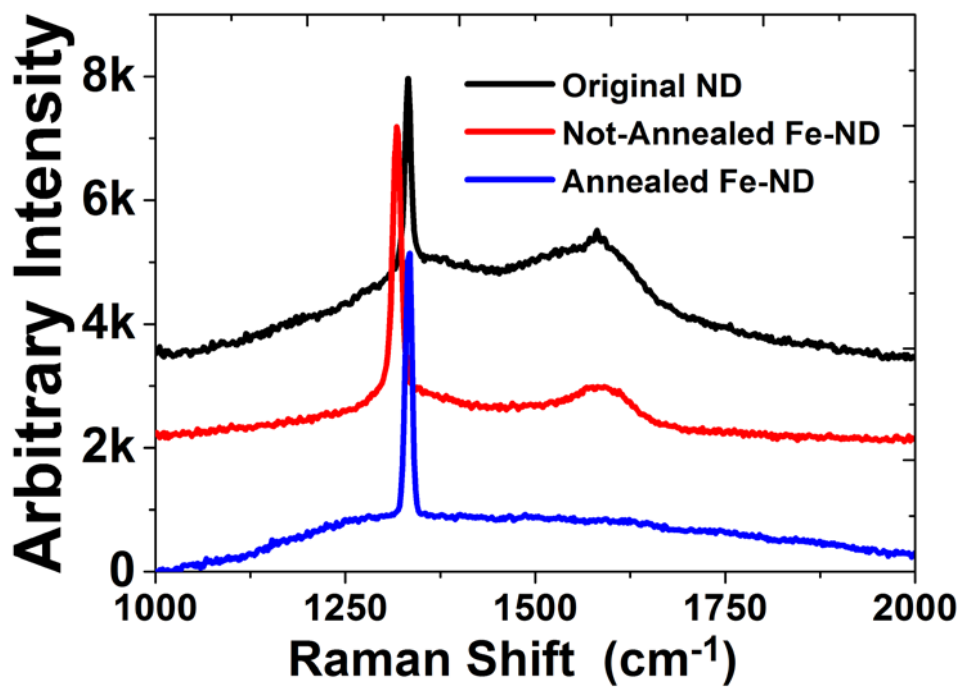

Figure 2. The figure shows measured Raman spectra of the original NDs, the NDs after Fe implantation and the NDs after implant and annealing. The original NDs had a diamond peak at 1332.5 cm$^{-1}$ and a broad peak at 1579 cm$^{-1}$, which is probably due to surface graphitic structure. After annealing, the diamond peak restored to the original position of 1332.4 cm$^{-1}$ and the FWHM was reduced to 11.1 cm$^{-1}$.

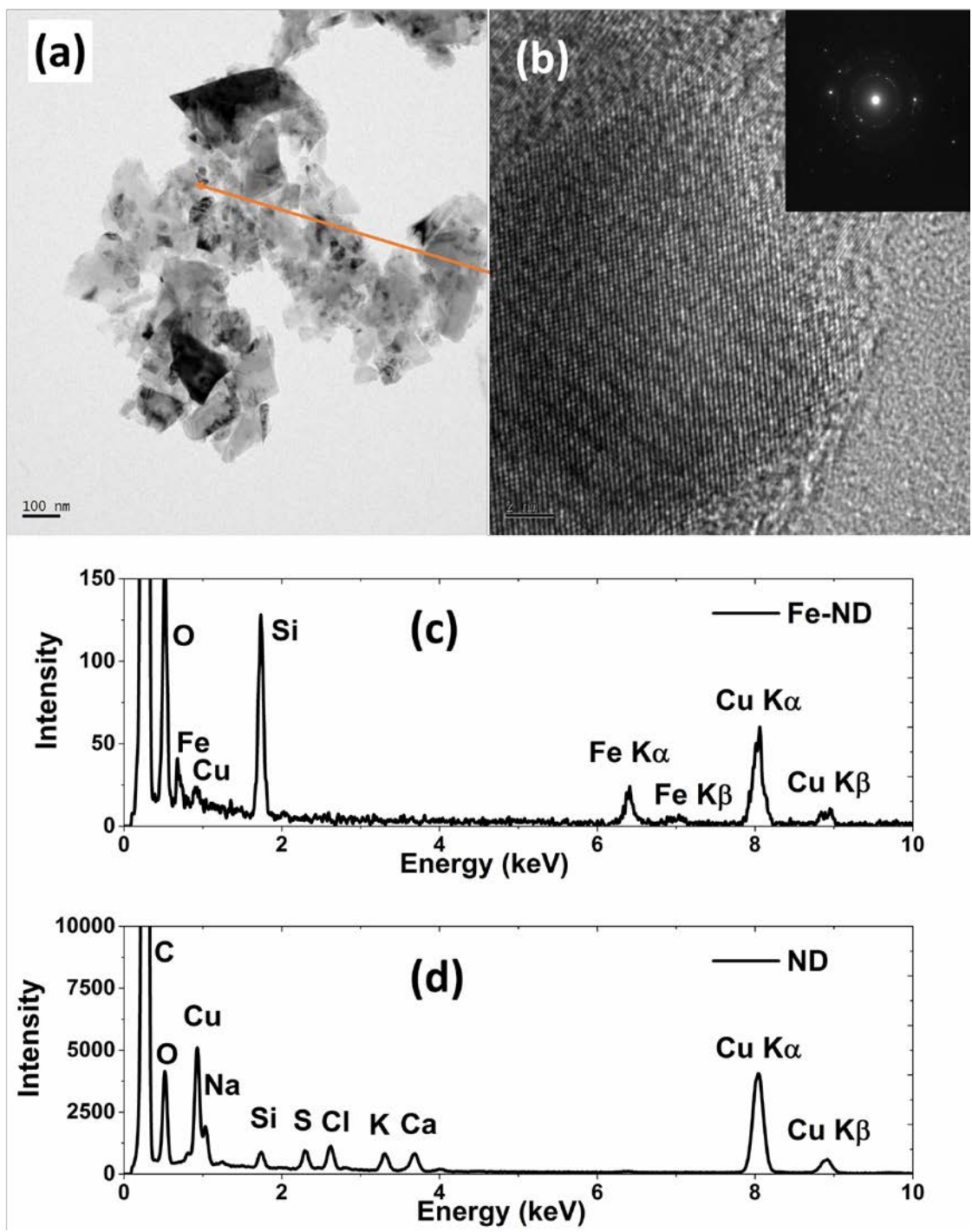

Figure 3. The aggregated NDs image is shown in (a) and that of a high resolution lattice image of a single ND is shown in (b). The spectrum of the Fe-implanted ND with the image shown in (b) is shown in (c) and the Fe content in the ND was estimated around 0.2% in weight. The EDX spectrum of an original ND is shown in (d).

Acknowledgements

We thank Prof. C.P. Lee of the department of electronics engineering, National Chiao Tung University, for his kind and valuable comprehensive discussion.